\documentclass[aps,pra,superscriptaddress,twocolumn]{revtex4-1}
\usepackage{bm} \usepackage{graphicx} \usepackage{amsmath}
\usepackage{amsthm,stmaryrd}
\usepackage{amssymb} \usepackage{epstopdf} \usepackage{txfonts}
\usepackage{color}
\usepackage{subfigure}

\newcommand{\ket}[1]{|{#1}\rangle}
\newcommand{\bra}[1]{\langle{#1}|}
\newcommand{\mean}[1]{\langle{#1}\rangle}

\DeclareRobustCommand\openzero{\leavevmode\hbox{0\kern-.55em0}}
\mathchardef\minus="002D

\newcommand{\impu}{\beta}

\begin{document}

\title{A Toolbox for Linear Optics in a 1D Lattice via Minimal Control}

\author{Enrico Compagno}
\affiliation{Department of Physics and Astronomy, University College London, Gower Street, WC1E 6BT London, United Kingdom}

\author{Leonardo Banchi}
\affiliation{Department of Physics and Astronomy, University College London, Gower Street, WC1E 6BT London, United Kingdom}

\author{Sougato Bose}
\affiliation{Department of Physics and Astronomy, University College London, Gower Street, WC1E 6BT London, United Kingdom}

\date{\today}
\begin{abstract}
Tight binding lattices offer a unique platform in which particles may be either
static or mobile depending on the potential barrier between the sites. How to
harness this mobility  in a many-site lattice for useful operations is still an
open question. We show how effective linear optics-like operations between
arbitrary lattice sites can be implemented by a minimal local control which
introduces a local impurity in the middle of the lattice. In particular we show
how striking is the difference of the two possible correlations with and without
the impurity. Our scheme enables the observation of the Hong-Ou-Mandel effect
between
distant wells without moving them next to each other with, {\it e.g.}, tweezers.
Moreover, we show that a tunable Mach-Zehnder interferometer is implemented
adding a step-like potential, and we prove the robustness of our linear optics
scheme to inter-particle interactions.
 
\end{abstract} 

\maketitle
Linear optical networks are indispensable tools for both fundamental investigations of quantum interference phenomena and for practical applications. Beam splitters  acting on two modes 
enable one to design simple two output interferometers such as the Mach-Zehnder and to observe
bosonic behavior of two incident particles in the most striking way through the
Hong-Ou-Mandel effect where the probability of one photon in each output is
completely suppressed. The same types of effects form the bedrock of linear
optical quantum computation \cite{knill_scheme_2001,popescu_knill-laflamme-milburn_2007}, and of the boson sampling device \cite{Sampling,SamplingExperimental,BSE2,BSE3,BSE4}. The recent atomic realization of a controlled beam splitter in a double well potential \cite{kaufman} highlights the importance of atomic linear optics. This, and the recent unprecedented abilities to initialize and measure the positions of individual atoms \cite{GreinerNew,fukuhara_microscopic_2013,fukuhara_quantum_2013,BoseNV}, raise the intriguing question: can we use a many-site lattice for performing arbitrary linear-optics operations? Large lattices are indeed required for many applications, such as boson sampling where the complexity increases dramatically when the number sites is much larger than the number of particles. 

At a first glance, the realization of arbitrary operations seems improbable, as atoms on a multi-site lattice typically perform a ``quantum walk'' which is dispersive. This severely limits the observability even of basic linear-optics effects, such as bosonic bunching and/or fermionic anti-bunching, as the particles quickly spread out between multiple modes \cite{omar_quantum_2006,GreinerNew,fukuhara_microscopic_2013,sansoni_two-particle_2012,crespi_anderson_2013,liu_quench_2014,Nikolo,qin_quantum_2014,lahini}.
In fact, such phenomena cannot be observed unless the particles are 
 nearest neighbors or in the same site
\cite{lahini}, even in the interacting case
\cite{GreinerNew,GreinerNew,fukuhara_microscopic_2013}. 
Obviously a new methodology is required 
in an atomic multi-site lattice
for neat two mode
demonstrations of such effects (as with two photons on a beam splitter 
\cite{hong_measurement_1987}  or matter waves 
\cite{lewis-swan_proposal_2014}). 

Motivated as above 
we show (i) how to implement {\it remote} linear optics via the dynamics 
of trapped neutral atoms interacting via the Bose-Hubbard Hamiltonian; 
(ii) how to improve the efficiency of our scheme by introducing a minimal engineering 
of the couplings.
Unlike other studies to simulate 
{\it specific} linear optical effects
\cite{kaufman,vaselago,kovachy}, our purpose is to convert the tight-binding 
lattice to a wire for \emph{scalable} and \emph{arbitrary}
linear optics transformation between static atoms in distant sites
(stationary and measurable both ``before" and ``after" the linear optics operation), 
with minimal control.

\begin{figure}[t]
  \centering
  \includegraphics[width=\columnwidth]{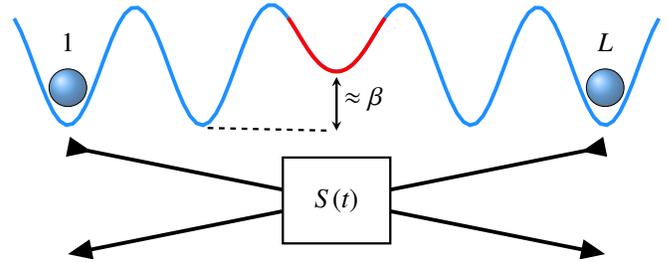}
  \caption{ { Scheme for remote linear optics in a multi-site open boundary lattice}. 
  Two particles initially in distant sites undergo an effective linear-optics
  transformation through scattering across an optical impurity. Free space
  evolution is replaced by ``quantum walk'' in a lattice. 
}
  \label{f.scheme}
\end{figure}

The first step to pursue this goal is the development of a scalable
procedure to realize a beam
splitter transformation between sites which are far from each other
(cf. Fig.\ref{f.scheme}). 
In our scheme, a {\it tunable}
remote beam splitter is realized by inserting a local
static defect at a single site in the lattice,  
as shown in Fig.~\ref{f.scheme}. This
enables us to study the two-mode Hong-Ou-Mandel effect,
as well as a Mach-Zehnder interferometer,
between atoms at {\it two distant} lattice sites.
Despite that dispersion limits the observability of linear optical effects in
our system for an unmodulated chain, for long chains we show quantitatively that
the efficiency of our scheme is close to one, once we introduce minimal
engineering schemes for the couplings, that are within the feasibility in current experiments.
On the other hand, for some transformations, a unit efficiency can be achieved
at the expense of a {\it full} engineering of the couplings \cite{PerfectBS}. 
Moreover, compared to Ref. \cite{PerfectBS} we also show how to introduce an
additional tunable phase factor suitable for interferometric applications. 
To appreciate the robustness of linear optics with interacting atoms we study the transition from bunching to anti-bunching as a function of the on-site interaction $U$ and we find that 
the critical value for the transition is curiously close to the superfluid-Mott critical point.

There are many schemes to implement quantum operations between two atoms
using an active transport of the particles  or an active
reshaping of the lattice.  For instance, 
quantum gates have been implemented with Rydberg atoms
\cite{wilk_entanglement_2010,isenhower_demonstration_2010} and
in double well systems \cite{Anderlini2007,DeChiaraDoubleWell2008}, using a combination
of suitably designed pulses and natural interactions, 
and with spin dependent optical lattices 
\cite{JackschCollision,MandelCollision} using controlled collisions. 
Controlled interactions, via Feshbach
pulses in optical lattices, have been proposed to perform operations 
between actively movable and static register atoms
\cite{CalarcoMarker,WeitenbergMarker2011}. 
On the other hand, our proposal exploits the natural atom dynamics in a time-independent
configuration, and therefore makes more straightforward its experimental 
realization within the current technology. 

\section{Remote Linear Optics via Quantum Walks \label{sec:BSUniform}}
As a paradigmatic model we use the single band Bose-Hubbard Hamiltonian which in one
dimension reads \cite{lewenstein_ultracold_2012}
\begin{align}
  H = -\sum_{j=1}^{L-1}\frac{J_j}{2} \left[a_j^\dagger a_{j+1} + {\rm h.c.} \right]
  + \sum_{j=1}^L\left[\frac{U_j}{2} \,n_j(n_j-1)-\mu_j \,n_j\right]~,
  \label{e.BH}
\end{align}
being $a_j^\dagger~(a_j)$ the boson creation (annihilation) operators, 
$n_j{=}a_j^\dagger a_j$ the number operator, $L$ the lattice sites,
$J_j$ the tunneling rate, $U_j$ the on-site interaction and $\mu_j$ the local
chemical potential.
Firstly we consider the case of a uniform chain $J_n{=}J$, $U_{n}{=}U$. In the fermionic case we model the system as in \eqref{e.BH} with the operators $a_j$ substituted by their fermionic counterparts and $U{=}0$. 
By varying the ratio $U/J$ the system undergoes a quantum phase transition 
to the Mott insulator phase where the number of particles per lattice site 
is fixed to a constant value that depends on the parameters $(J,U,\mu)$
\cite{kuhner_phases_1998,rapsch_density_1999}. 

A physical realization of the tight-binding model \eqref{e.BH} is with cold
atoms in optical lattices both for bosons
\cite{jaksch_cold_1998,greiner_quantum_2002}, hard-core bosons
\cite{rigol_universal_2004} and fermions
\cite{jordens_mott_2008,modugno_production_2003}.  Alternative implementations
are systems of interacting polaritons \cite{hartmann2006strongly}, coupled
quantum dots \cite{nikolopoulos2004electron} and photonic lattices
\cite{bellec2012faithful}. The optical lattice implementation has in particular
several appealing features, because it represents a good balance between
insulation from the environment and excellent controllability. The site
dependent coupling constants in \eqref{e.BH} can be tuned via addressable
optical lattices \cite{wang_fault-tolerant_2013}, created projecting an electric
field profile via holographic masks \cite{bakr_quantum_2009} or via micro-mirror
devices \cite{GreinerNew}. The system exhibits excellent coherence properties
due to the weak coupling with the environment, whereas   decoherence effects are
mainly due to spontaneous emission \cite{pichler_nonequilibrium_2010}.  These can be strongly suppressed when a blue-detuned light is employed to create the lattice, allowing a quasi-unitary particle dynamics \cite{pichler_nonequilibrium_2010,Sakar}.

Finally the amount of control available in current experiments allows single
particle initialization and read-out via single-atom addressing techniques and
fluorescence imaging microscopy
\cite{weitenberg_single-spin_2011,wurtz_experimental_2009,bakr_quantum_2009,BlochNew}.
Once in the Mott insulating phase, the system can be initialized via single site
addressing in a state with a few localised particles, and the dynamics of a
single traveling particle \cite{fukuhara_quantum_2013} and of two (interacting)
particles \cite{fukuhara_microscopic_2013,GreinerNew} can be observed. State
preparation fidelity is around 98\% while single atom detection is possible with
efficiency around 99\% \cite{BlochNew,schreiber_observation_2015}. Due to
light-assisted collisions and pairwise atom loss in fluorescence imaging, pairs
of atoms in the same site are detected applying a magnetic gradient before the
fluorescence detection technique \cite{GreinerNew}. Some progresses have been made recently via occupation-dependent interplane transport \cite{GreinerNewII}.

All linear optics operations can be performed with beam splitters, phase shifters and mirrors \cite{reck_experimental_1994}. We first focus on implementing tunable beam-splitter operations between distant sites in an optical lattice  via the scheme displayed in Fig.~\ref{f.scheme}. 
We consider an odd chain ($L{=}2N{+}1$) with a local potential on site $N{+}1$ which gives rise to an impurity in the chemical potential \cite{jaksch_cold_1998}:
$\mu_j {=} \mu {+} J\impu\,\delta_{N+1,j}$. 
Once the particle number is fixed, the constant term $\mu$ only produces 
an irrelevant global phase. 
On the other hand, 
the potential barrier $\impu$ favors the splitting 
of an incoming particle into a transmitted and a reflected component. 
We set the initial position of the particle on site 1 and we
define the transmission and reflection coefficients ($T$ and $R$ respectively) as 
$  T(t) {=} \bra0 a_L\, e^{-i t H} \,a_1^\dagger \ket 0$, $
R(t) {=} \bra0 a_1\, e^{-i t H} \,a_1^\dagger \ket 0$,
being $\ket 0$ the vacuum state. $R(t)$ represents the probability amplitude
that the particle returns to site 1 after time $t$, while
$T(t)$ is the probability amplitude to reach
the opposite end (site $L$) on time $t$. 
Due to the symmetries of the system, the same coefficients also describe 
the case of a particle initially located on the end site $L$. 
These initial locations are chosen to force the particles to
travel on a single direction, namely towards the optical impurity 
and ultimately towards the other end. We find analytical expressions for $T(t)$
and $R(t)$ by using a technique for computing eigenvalues and eigenvectors of a
quasi-uniform tridiagonal matrix \cite{banchi_spectral_2013}. Details of the calculations are reported in appendix \ref{app:OddChain}. For the relevant values of $\impu$ the coefficients $T(t)$ and/or $R(t)$ display their first maximum at the same time (hereafter named $t^*$), which do not depend on $\impu$. Therefore $t^*$ coincides with the transmission time of 
the $\impu{=}0$ case ($t^*{\approx}N J^{-1}$ with some finite size corrections \cite{banchi_nonperturbative_2011}). 

\begin{figure}[t]
\centering
\subfigure{\includegraphics[width=\columnwidth]{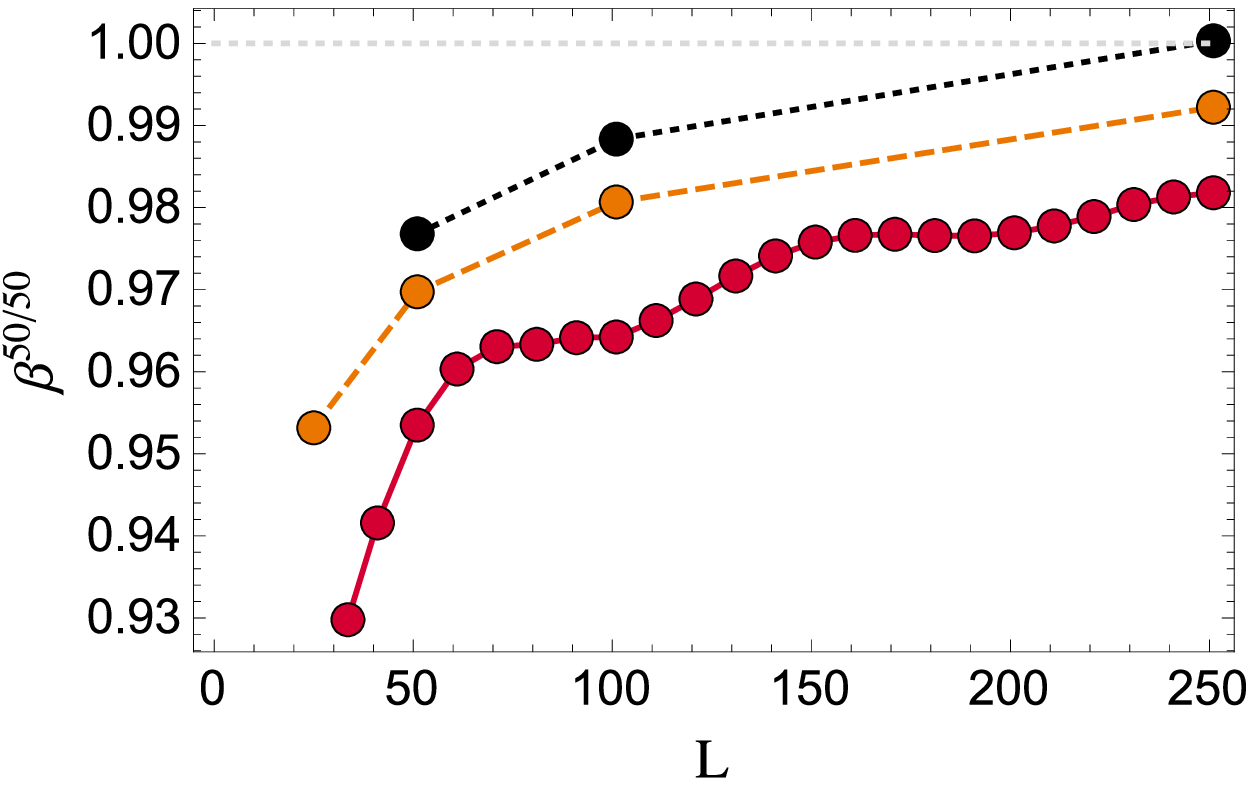}}
\subfigure{\includegraphics[width=\columnwidth]{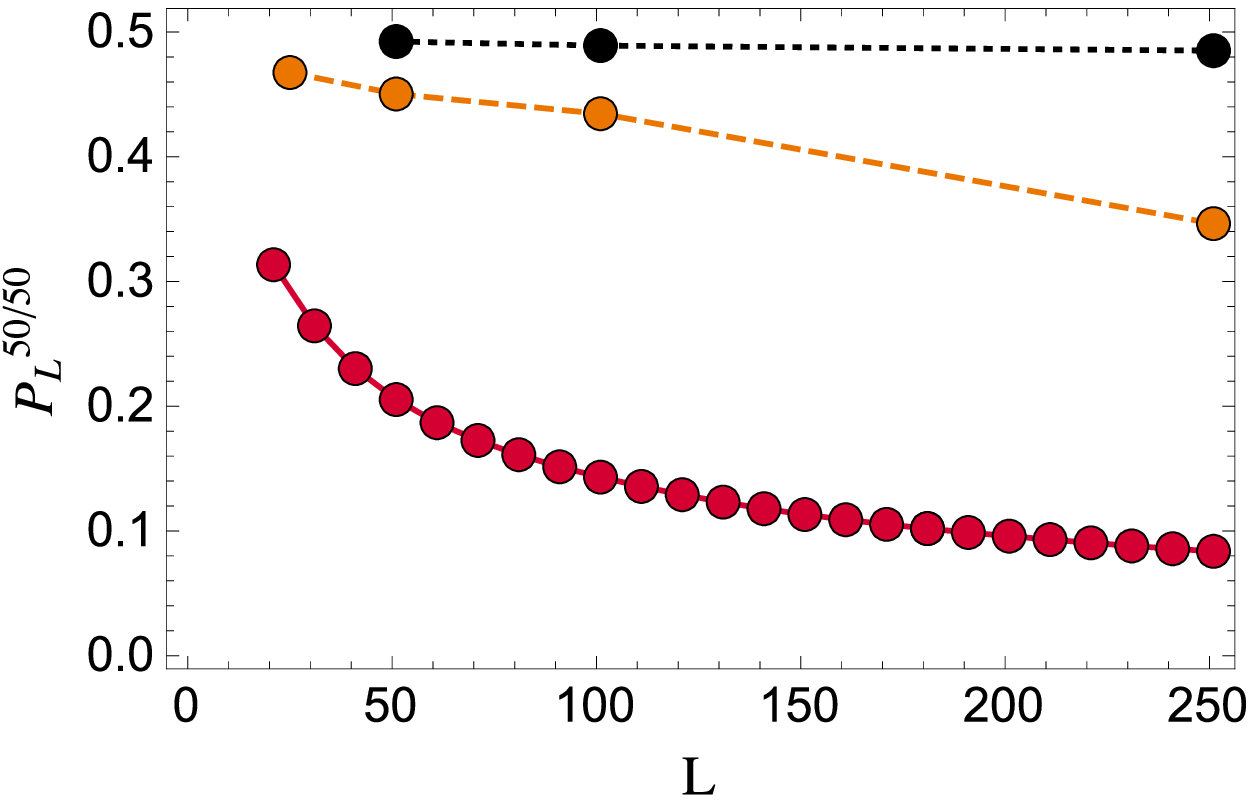}}
\caption{(color online) 
(top) Optimal value of $\beta$ for the $50{/}50$ beam splitting effect as
function of the chain length for (red) uniform couplings, (orange) optimal
couplings and (black) the double optimal coupling scheme. The grey dotted line is the
asymptotic value $\beta^{50/50}{=}1$. (bottom) Output probability
$P_{L}^{50/50}{=}\vert T(t^*)\vert^2$ as a function of the chain length $L$ for
$\ket{\psi(t{=}0}{=}a^{\dag}\ket{0}$, when $\beta{=}\beta^{50/50}$.}
\label{fig:B5050}
\end{figure}

Via the coefficients $R(t^*)$ and $T(t^*)$ we define an effective beam splitter operator
whose input ports are sites 1 and $L$ at time $t{=}0$, and whose output ports are 
the same sites at time $t^*$:
\begin{align}
  S(t^*) = \begin{pmatrix}
    R(t^*) & T(t^*) \\ T(t^*) & R(t^*)
  \end{pmatrix} 
  \approx D\begin{pmatrix}
    \frac{\impu}{i+\impu} &
    \frac{-i}{i+\impu} \\
    \frac{-i}{i+\impu} &
    \frac{\impu}{i+\impu} 
  \end{pmatrix}+\mathcal O(L^{-1})~,
  \label{e.S}
\end{align}
where the second equality holds for $L{\gg}1$. 
Details of the calculations are in Appendix A. 
The effective beam splitter operator Eq.\eqref{e.S} is the product of 
a damping factor $D{=}\mathcal O(L^{{-}1/3})$  and a unitary matrix
$\tilde S{=}S(t^*)/D$. The damping factor is due to the non-linear dispersion
relation of the model: the wavefunction is not perfectly reconstructed 
at $t^*$ and there is some probability to find the particle far from the ends.
However the factor $|D|$ can be made arbitrary close to one with a further engineering of the couplings that avoids wave-packet dispersion. We specifically address this point in the following section. Eq.~\eqref{e.S} quantifies the splitting of traveling particles into transmitted and reflected components. For $\impu{\to}0$ there is just
the transmitted component, whereas for $\impu{\to}\infty$ only the reflected 
component is non-zero; a 50/50 beam splitter is implemented when $\impu{=}\impu^{50/50}{=}1$. For finite $L$ there is a $\mathcal O(L^{ {-}2/3})$ correction 
to the value of $\impu^{50/50}$ which is therefore obtained numerically, via
exact diagonalization methods, by imposing $\vert R(t^*)\vert {=}\vert T(t^*)\vert $. 
\begin{figure}[t]
\centering
\includegraphics[width=\columnwidth]{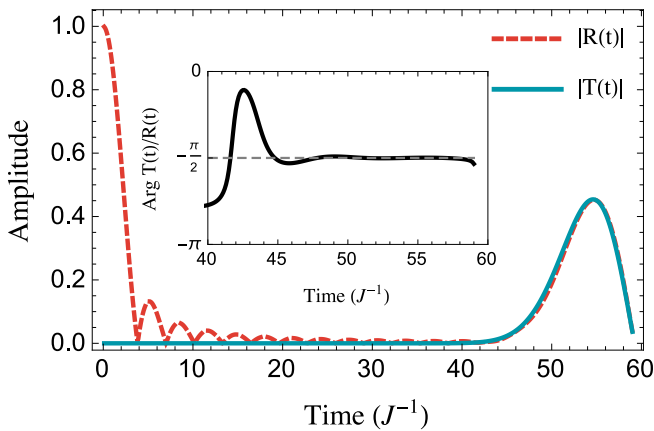}
\caption{(color online) { Reflection and transmission coefficients for a single particle.}
Amplitudes $R(t)$ and $T(t)$  and relative phase (inset) 
  as a function of time for a uniform chain with $L{=}51$ and 
$\impu{=}\impu^{50{/}50}{\simeq}0.95$ in the initial state $\ket{\psi(t{=}0)}{=}a^{\dag}\ket{0}$. The transfer time is $t^*{\simeq}55/J$. }
\label{f.RT}
\end{figure}
The dynamics of $R(t)$ and $T(t)$ in the 50/50 regime is shown in
Fig.~\ref{f.RT} when $L{=}51$. In Fig. \ref{fig:B5050} we report the results
obtained for the $\beta^{50{/}50}$ values and for the factor $D$ as a function
of the chain length $L$ for the uniform chain (red points). The damping factor
is investigated via the output probability $P_{L}^{50/50}{=}\vert T(t^*)\vert^2$
as a function of the chain length $L$, for a single particle in the initial state $\ket{\psi(0)}{=}a_1^\dag \ket{0}$. 
We note that 
particle dispersion limits the observability of the beam splitter effects in
long chains. In the following section we analyze how an extra minimal
engineering of the couplings can improve the efficiency of our scheme.

\subsection{Efficiency improvement via engineered coupling schemes}
Wave-packet dispersion during the hopping dynamics can be drastically reduced by
engineering the couplings of the lattice. 
By using a full-engineering of the interactions one can obtain a perfect
transmission to arbitrary distant sites
\cite{kay_perfect_2010,Jex,christandl_perfect_2004,PerfectBS,clark_efficient_2005}.
However, 
the full-engineering could be too demanding \cite{Sousa2014} in comparison with the level of fidelity required for the implementation of most quantum information processing tasks. 
In fact, for many practical purposes, an almost perfect transfer, achieved with
a minimal engineering
\cite{OptimalTransfer,DoubleOptimalTransfer}, provides a high enough fidelity without
requiring a fine tuning of many parameters.

Motived by the pursuit of minimal control strategies, we consider two optimal
transfer engineering schemes, that require respectively the control of the first and the last 
tunneling couplings \cite{OptimalTransfer}, or the control of the first two and last
two couplings \cite{DoubleOptimalTransfer}. We refer to the first
coupling patterns as the ``optimal couplings'' while we call the latter one as the
``double optimal coupling'', and we compare them with the uniform coupling case.
As for the uniform case the $\beta$ impurity represents a local perturbation
whose effect is to split the incoming particle wavefunction
$\ket{\psi(t{=}0)}{=}a_1^\dag\ket{0}$ and to produce a beam splitting effect. In Fig. \ref{fig:B5050}  
we study as a function of the chain length $L$ both the value of $\beta$ that
fulfills the $50{/}50$ condition, and 
the output probability $P_{L}^{50/50}{=}\vert T(t^*)\vert^2$, whose  
deviation from the ideal case ($P_{L}^{50/50}{=}0.5$) shows the effect of the damping factor $D$. 
 We observe that, for fixed $L$, the impurity strength $\beta$ in the optimal
 coupling schemes is closer  to the  asymptotic value $\beta^{50{/}50}{=}1$,
 compared to the uniform case. 
 Moreover, from the analysis of $P_{L}^{50/50}$, 
 we observe that the optimal coupling schemes, in particularly the double
 optimal one, offer a remarkable improvement of the transmission quality
 compared to the uniform case, and enable one to obtain  an almost
ideal beam splitting behavior.  This is of fundamental importance for the experimental realization and for technological applications.

\subsection{Even Chain}
\begin{figure}[h]
\subfigure{\includegraphics[width=\columnwidth]{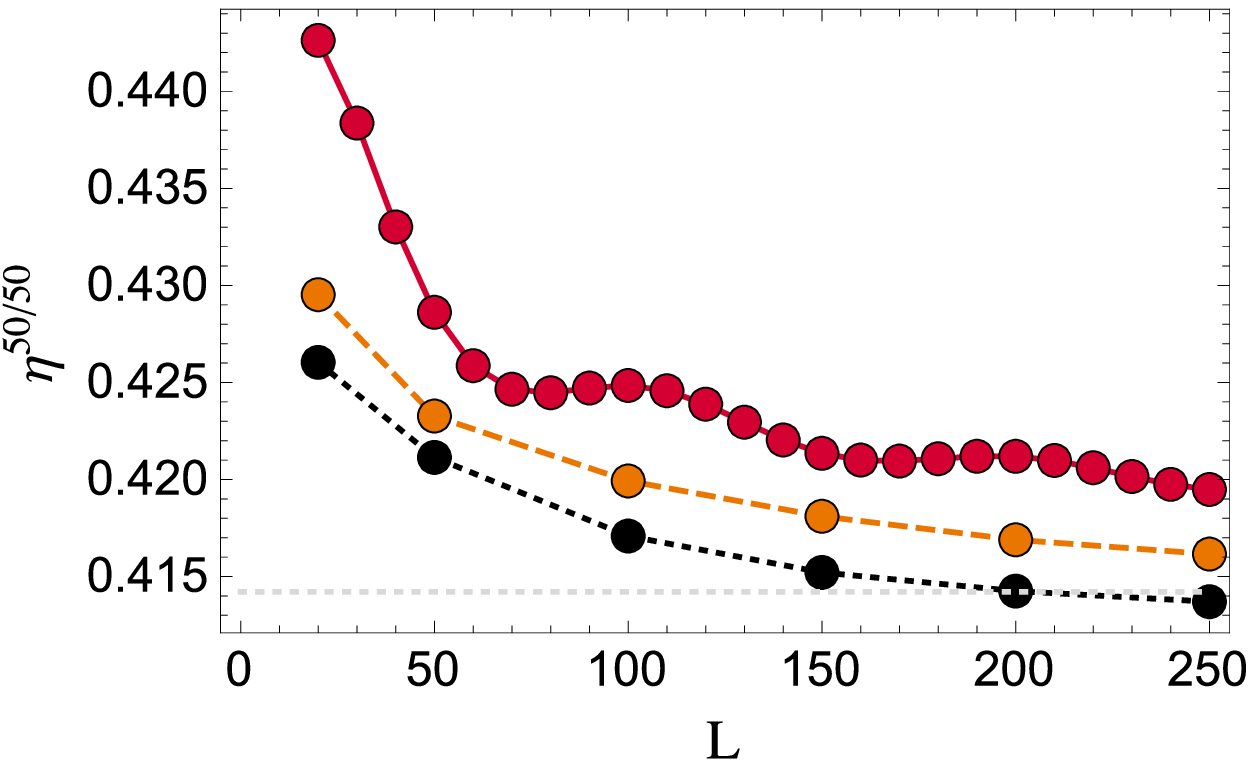}}
\subfigure{\includegraphics[width=\columnwidth]{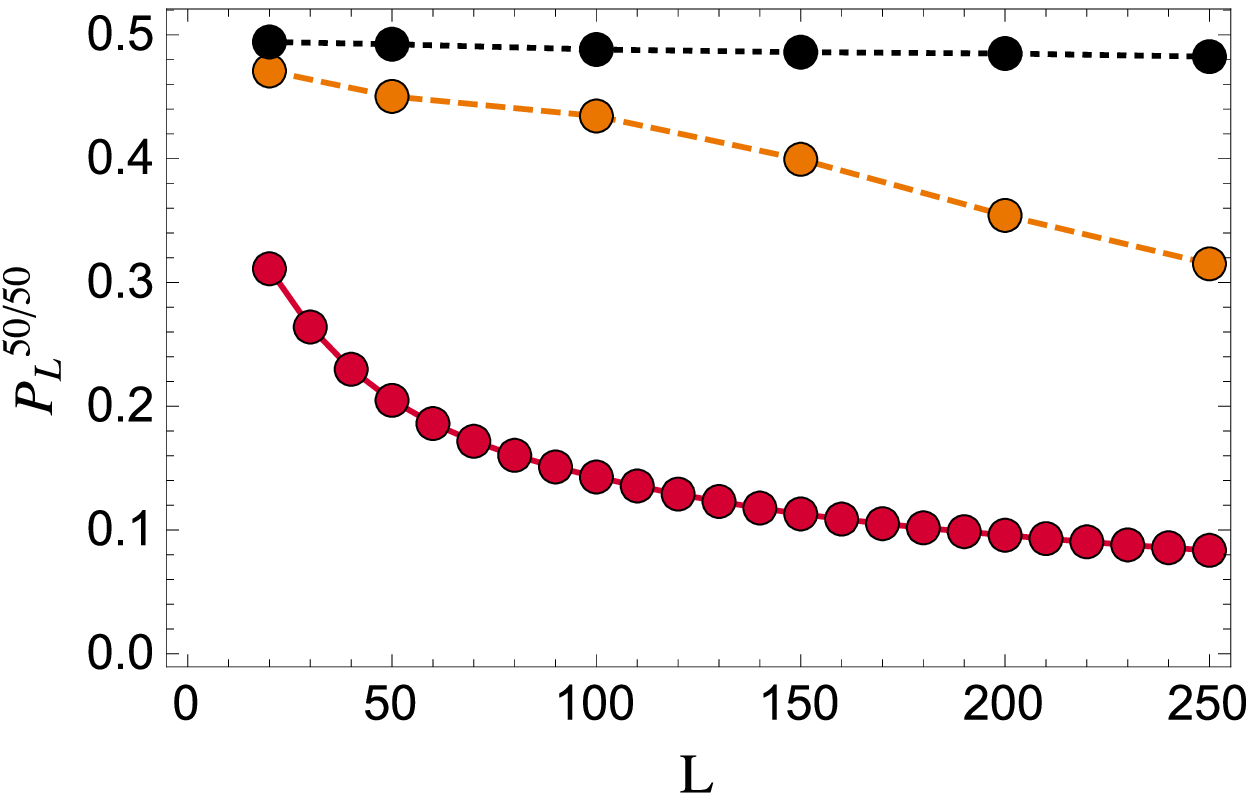}}
\caption{(color online) 
(top) Optimal value of $\eta$ for the $50{/}50$ beam splitting effect as
function of the chain length for (red) uniform coupling scheme, (orange) optimal
couplings, and (black) double optimal couplings.  The grey constant line
represents the asymptotic value for $L{\gg}1$, namely $\eta{=}\sqrt{2}{-}1$.
(bottom) Output probability $P_{L}^{50/50}{=}\vert T(t^*)\vert^2$ as a function of
the chain length $L$ for $\ket{\psi(t{=}0}{=}a^{\dag}\ket{0}$, when
$\eta{=}\eta^{50/50}$. }
\label{fig:EvenUniform}
\end{figure}

It is worth mentioning that the previous scheme, valid for odd chains, can be
applied to even chains by replacing the impurity in the chemical potential with
a coupling impurity $J'$ in the middle of the chain. 
 As for the odd chain case, the Hamiltonian in the single particle sector is a
 quasi-uniform tridiagonal matrix. This allows us to evaluate how the reflection
 and the transmission coefficients depend on the impurity strength $\eta{=}J'/J$ in the limit of $L{\gg}1$. The effective beam splitter operator $S(t^*)$ is evaluated in Appendix \ref{sec:EvenChain} and it is
\begin{align}
S(t^*) = 
\begin{pmatrix}
R(t^*) & T(t^*) \\ 
T(t^*) & R(t^*)
\end{pmatrix} 
\approx D 
\begin{pmatrix}
\frac{ 1-\eta^2}{1+\eta^2} &  \frac{-2 i \eta}{1+\eta^2} \\
\frac{-2 i \eta}{1+\eta^2} & \frac{ 1-\eta^2}{1+\eta^2}
\end{pmatrix}
+\mathcal O(L^{-1})~,
\label{eq:oddtransbs}
\end{align}
From the latter we find that a 50/50 beam splitter is obtained when
$\eta{=}\sqrt{2}{-}1$. Deviations from this value due to finite size effects have
been investigated, as for the odd chain, by a numerical  minimization of the
difference between the reflection and the transmission coefficients. In Fig.
\ref{fig:EvenUniform} we plot the obtained optimal strength $\eta^{50/50}$ as
function of the chain length $L$ and  the resulting output probability
$P_{L}^{50/50}{=}\vert T(t^*)\vert^2$ at the transmission time, using different
engineering schemes. Qualitatively, the results obtained for even chains are
comparable with those found in the odd chain case. 

\section{Long distance Hong-Ou-Mandel interference} 
\begin{figure*}[t!]
  \centering
  \includegraphics[width=1\textwidth]{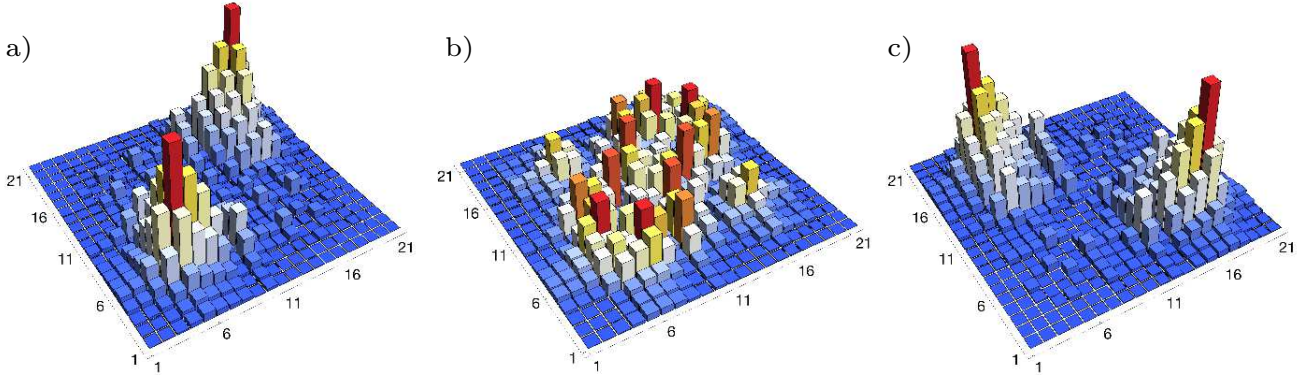}
  \caption{(color online) {Two particle correlation, $P_{jk}/ P_{jk}^{\rm
  max}$}, for a unmodulated chain ($J_n{=}J$) with Hamiltonian \eqref{e.BH},
  with an impurity in the middle of the chain
  $\impu{=}\impu^{50/50}{\simeq}0.94$, for bosons (a),  fermions/hard-core bosons (c) and for the intermediate regime $U/J{\simeq}0.71$ (b). It is $L{=}21$, $t{=}18J^{-1}$. To cancel possible boundary effects,  $P_{jk}(t)$ is plotted for $t{\simeq}0.75t^*$.}
\label{f.P}
\end{figure*}

\begin{figure}[htb]
  \centering
  \includegraphics[width=1\columnwidth]{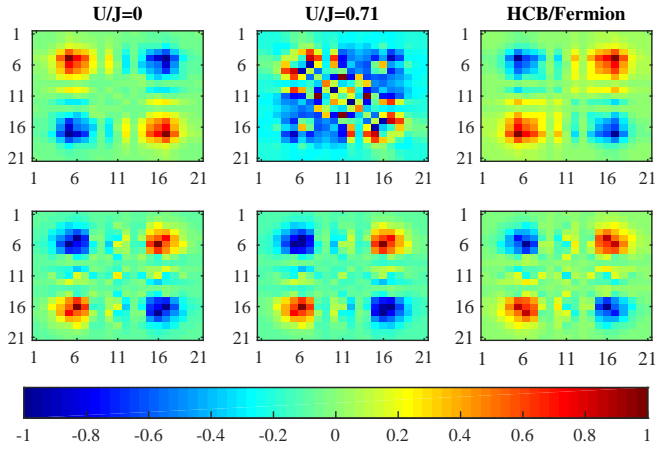}
  \caption{(color online) (top) Two particle correlations $C_{jk}{=}P_{jk}{-}P_j
  P_k$ for a unmodulated chain ($J_n{=}J$) with Hamiltonian \eqref{e.BH}, with an impurity in the middle of the chain $\impu{=}\impu^{50/50}{\simeq}0.94$. Here $L{=}21$ at time $t{\simeq}18 J^{-1}$, for (left) bosons, (right) fermions/hard-core bosons and for the intermediate regime (center) $U{/}J{=}0.71$. (bottom) Correlation functions $C_{jk}$ in absence of impurity, $\beta{=}0$.}
\label{fig:CorrFuncFukuhara}
\end{figure}
The possibility to generate an effective beam splitter transformation between
distant sites opens up to generate multi-particle interference effects. Peculiar
quantum statistical effects appear when there are two traveling particles
initially located at opposite boundary sites $1,L$. When $U{\to}\infty$ the
Bose-Hubbard Hamiltonian \eqref{e.BH} describes hard-core bosons and the system
is equivalent, via the Jordan-Wigner transformation, to a spin-$\frac 12$ chain with XX interactions \cite{lewenstein_ultracold_2012}. In our simulations, the dynamics of {\it hard-core} bosons and fermions are indistinguishable. The evolution is  described in the Schr\"odinger  picture by
the state $\ket{\Psi(t)} {=} \sum_{jk} A_{jk}^{b/f}(t) a_j^\dagger a_k^\dagger \ket 0$, where the subscript b/ f explicitly denotes the bosonic/fermionic case. Here $A_{jk}^{b}(t){=}A_{jk}(t){/}\sqrt{1+\delta_{jk}}$ for two bosons, while $A_{jk}^{f}(t){=}A_{jk}(t)$ for two fermions. The amplitudes $A_{jk}(t)$ evolve through the Schr\"odinger equation as $i\dot A_{jk}(t) {=} \sum_{mn} K_{jk,mn} A_{mn}(t)$, where $A_{jk}(0){=}\delta_{j1}\delta_{kL}$ and $K_{jk,mn}$ is obtained from \eqref{e.BH} via the algebra of commutation relations and $\hbar{\equiv}1$. We set $A_{jk}{=}0$ for $j{>}k$ ($j{\ge}k$) for bosons (fermions).

The two-particle interference effects are studied via the correlator
$P_{jk}^{b/f}(t){=}\bra{\Psi(t)} a_j^\dagger a_k^\dagger a_k a_j\ket{\Psi(t)}$
between  two sites $j$,$k$ ($j{\le}k$) \cite{lahini2009} that in turn can be
expressed in terms of probability amplitudes $A_{jk}(t)$ by using the Wick theorem. When $U{=}0$ the model is quasi-free and $P^{b/f}_{1L}(t){=}|T^2(t){\pm}R^2(t)|^2$, $P^b_{11}(t){=}P^b_{LL}(t){=}4|T(t)R(t)|^2$.
This quantity is experimentally accessible in an optical lattice implementation
\cite{GreinerNew,fukuhara_microscopic_2013,fukuhara_quantum_2013}. In the
specific, in Ref.\cite{GreinerNew} single particles are initialized in the
lattice and pairs of atoms in the same site are separated, with a magnetic
gradient, prior to the detection to avoid light-assisted collision processes. 

Owing to the explicit expression of the effective beam splitter
operator \eqref{e.S}, one obtains perfect bunching (anti-bunching) with bosons
(fermions) at time $t^*$ when $\impu{=}\impu^{50/50}$. In Fig.~\ref{f.P} we show
the correlator $P_{ij}$ for two particle initially in $\ket{\psi(0)}{=}a_1^\dag
a^\dag_L \ket{0}$ for a chain with $L{=}21$ at $t{=}18 J^{-1}$ (before $t^*$ to
avoid boundary effects) in the $50{/}50$ beam splitter configuration, namely
with $\beta{=}0.94$. Although the effect is damped in a homogeneous chain
because of the dispersive transmission (i.e. $|D|{\neq}1$), for bosons we
clearly observe an
increase of the probability to have two particles on the same side of the chain,
while fermions show a maximum of $P_{jk}$ when the two particles are in opposite
sides. In Fig. \ref{fig:CorrFuncFukuhara} (top) the bunching/antibunching behavior is evaluated by studying the correlation function $C_{jk}{=}P_{jk}{-}P_j P_k$, where $P_j{=}\sum_k P_{jk}$ takes into account the distinguishable motion of the atoms in the lattice, as in \cite{fukuhara_microscopic_2013,lahini}. As shown in Fig.~\ref{f.P} and in Fig.~\ref{fig:CorrFuncFukuhara}, bosons bunch while fermions and hard-core bosons anti-bunch. 

One important question to address is whether any bunching effect is observable
without the impurity $\beta{=}0$ in the Hamiltonian \eqref{e.BH}. 
Hanbury Brown-Twiss correlations have been observed in an optical lattice when
the two particles are initially in neighboring sites \cite{GreinerNew}. 
On the other hand, when the particles are initially in distant sites 
 no bunching effect appears even when the onsite interaction $U{/}J$ is not zero.
This is shown in 
the lower panel of Fig.  \ref{fig:CorrFuncFukuhara}, where  the correlation function
$C_{jk}$ between two particles initially in sites $1$ and $L$ is plotted when
 $\beta{=}0$. 
A similar conclusion was given in Ref.\cite{lahini} where two particles initially
separated by an empty site were analyzed.
These results justify the importance of introducing an optical impurity to
generate an effective beam splitter transformation for incoming wavepackets, and
then to  produce interference phenomena such as the Hong-Ou-Mandel effect between distant sites.

\begin{figure}[t]
\centering
  \includegraphics[width=1\columnwidth]{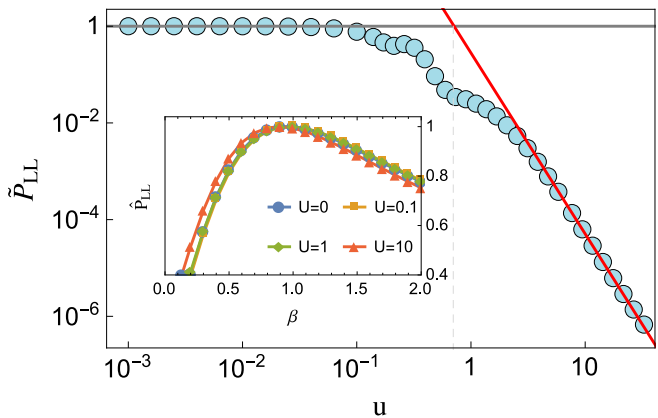}
  \caption{ (color online) {Transition from bunching to anti-bunching}. 
   Two-particle probability $\tilde P_{LL} {=}P_{LL}/P_{LL}^{U{=}0}$, in a uniform chain, to have two bosons in the last site as a function of $u{=}U/J$ for $\beta{=}\beta^{\rm opt.}$ and $t{=}t^*$, normalized with respect to the $U{=}0$ case for $L{=}51$. 
   (Inset) $\hat P_{LL} {=} P_{LL}/\max_\beta P_{LL}$ as a function of $\beta$ for fixed 
 values of $u$ ($L{=}51$). 
}
\label{fig:AnnU} 
\end{figure}

A finite $U$ reduces the double occupancy probability for two bosons on the same
site and one expects a transition from bunching to anti-bunching when $U$ goes
from 0 to $\infty$. Our results show that  $P_{jk}$ is almost indistinguishable
from the fermion case when $U {\gtrsim} 10J$. To highlight the transition from bosons to hard-core bosons in Fig.~\ref{fig:AnnU} we report the two particle correlator $P_{LL}(t^*,\beta^{\rm opt})$ for different $U$ and $L{=}51$, in a uniform chain, normalized with respect to the $U{=}0$ case. For fixed $U$, $t^*$ and $\beta^{\rm opt}$ are found numerically by maximizing $P_{LL}(t,\impu)$ around $t{\sim}L/J$ and $\impu{\sim}1$. As shown in the inset of Fig.~\ref{fig:AnnU}, the observed optimal $\impu$ for different $U$ is approximately equal to the value $\impu^{\rm 50/50}$ obtained when $U{=}0$. 
In fact, in the inset of Fig. \ref{fig:AnnU} we plot the two-particle
correlation $P_{LL}(t^*)$ as a function of the impurity strength $\beta$, and we
can clearly see that the $\beta$ which maximizes ${P}_{LL}$ is almost
independent on $U$. 
For $U {\lesssim} 0.1J$ the bunching effect is almost unaffected, while for $U {\gtrsim} 3J$ a power law decay occurs and there is a rapid change of behavior near $U{\simeq} J$. The threshold value $U^c$ separating the two regions is obtained by fitting the data in the power law region and taking the 
intersection value with the unit constant line. For $L{=}51$ in particular we estimate $U^c{\simeq} 0.71J$. The estimated $U^c$ is surprisingly similar to the critical value of the Mott insulator transition at the boundaries of a one dimensional chain (coordination number $z{=}1$) \cite{danshita_superfluid--mott-insulator_2011,PhDThesis:Greiner}. 
This numerical evidence raises interesting questions on the possibility to 
detect the Mott phase transition via specific features of the chain's
non-equilibrium evolution. 

\begin{figure}[t]
\centering
\includegraphics[width=1\columnwidth]{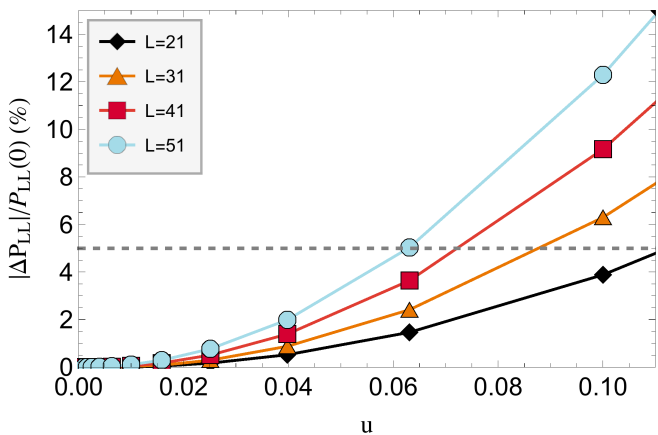}
\caption{(color online)  Relative variation of the  two-particle correlation
  $\vert \Delta P_{LL}\vert {/} P_{LL}(0)$ as a function of the on-site
  interaction $u{=}U{/}J$ in the weakly interacting regime $u{<}1$ for a uniform chain with length $L$. The gray dashed line represents a deviation of the $5\%$ with respect of the free-bosons case. }
\label{fig:UEffWeakCoupl}
\end{figure}

It is worth mentioning  that the weakly interacting regime $U{/}J{<}1$ is
realized in \cite{GreinerNew}, but also that the non-interacting regime $U{=}0$
has been experimentally achieved with Cs atoms loaded in a one-dimensional lattice, 
exploiting Feshbach resonances \cite{Meinert2014}.
In view of other possible experimental applications with different
atoms, 
we study in more detail how the bunching effect is affected by the
interaction. We evaluate the relative variation of the two particle correlator
$\vert \Delta P_{LL}\vert {/}P_{LL}(0)$, where $\Delta P_{LL}{\equiv}\vert
P_{LL}(u){-}P_{LL}(0)\vert$, as a function of the on-site interaction $u{=}U{/}J$
for a uniform chain with different chain lengths $L$. As it is shown in Fig.
\ref{fig:UEffWeakCoupl} the relative variation is lower than 5\% in the
range $0.06{<}u{<}0.1$, and the threshold  depends on the chain length. We argue
that this effect is due to the larger spatial extent of the two traveling
wavepackets ($\approx L^{1/3}$) 
that, accordingly, interact for more time in longer chains.

\section{Mach-Zehnder interferometer}
All the three optical elements which form a Mach-Zehnder 
interferometer can be obtained in a lattice: beam splitters have already been
discussed; mirrors are implemented by boundary
reflections and a phase shift can be obtained by freezing the hopping
so that the system acquires a phase because of the chemical potential. 
Alternatively a phase shift can be obtained by applying a localized field in a site different from the middle one. 
However, this does not allow for a continuous control on the generated phase factor 
and has the further disadvantage that, in this case, the reflected and the
transmitted components reach the two edges at different times. 

In view of experimental applications, it is more compelling to 
devise a scheme which minimizes the dynamical control required on the chain.
We show that the combined action of a beam splitter and a local phase-shifter 
can be achieved by applying a different optical potential to the right half of the chain: 
$\mu_j{=}J\gamma_R$ for $j{>}N{+}1$, being again
$\mu_{N{+}1}{=}J\beta^{50/50}$, and $\mu_{j}{=}0$, $j{\le}N$.
When $\gamma_R{\neq} 0$ a particle acquires a tunable 
phase $\phi{\propto}\gamma_R$ while 
it travels in the right part of the chain.
We find that an effective 50/50 beam splitter
can be obtained also when $\gamma_R{\neq}0$ by tuning 
$\impu^{50/50}_{\gamma_R}{\simeq} \beta_{\gamma_R{=}0}^{50/50}{\simeq} 1$.
As in the $\gamma_R{=}0$ case, 
the scattering matrix can be approximately factorized as
$S_\phi(t^*){\simeq}D\tilde
S_\phi(t^*)$, being $D$ a damping factor and $\tilde S_\phi$ a unitary matrix. 
When $\impu{=}\impu^{50/50}_{\gamma_R}$ the latter is found to be 
$\tilde S_\phi{\simeq}\frac1{\sqrt 2}\begin{pmatrix}
  1 & -ie^{i\phi} \\ -ie^{i\phi} & e^{2 i \phi}
\end{pmatrix}~$ where $\phi{=}\gamma_R t^*/\pi$ namely 
$\tilde S_\phi$ is a composition of a 50/50 beam splitter and two phase shifts. 
By applying the unitary matrix $\tilde S_\phi$ twice one obtains an optical
operation which is equivalent (up to local phase shifts) to a 
Mach-Zehnder unitary operation \cite{haroche_exploring_2013} 
(with phase $2\phi$). 

As an interesting application, we show that 
by properly tuning the phase $\phi{\propto}\gamma_R$ one can vary
the probability outcomes on the two output ports: as in a
Mach-Zehnder interferometer \cite{haroche_exploring_2013},
depending on $\phi$ a particle traveling from site 1 goes to
site $1$ or $L$ after time $2t^*$. 
\begin{figure}[t]
  \centering
    \includegraphics[width=1\columnwidth]{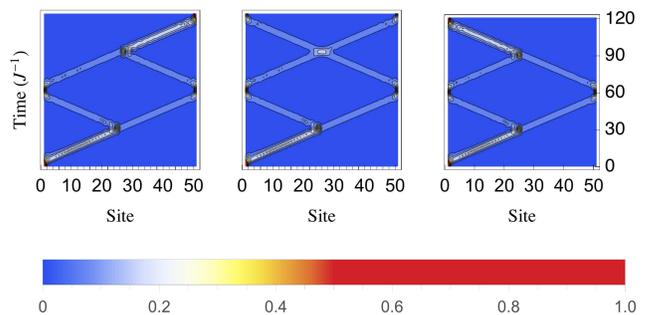}
  \caption{ Effective Mach-Zehnder interferometer where
    ${\impu}{=}{\impu}^{50/50}$ and $\gamma_R$ is tuned such that $\phi{=}0$
    (left), $\phi{=}\pi/4$ (centre) and $\phi{=}\pi/2$ (right) -- see the
    discussion in the main text. We plot $\mean{n_j(t)}$ vs position and time in
    a chain with $L{=}51$. The first two and the last two couplings are engineered according to  \cite{DoubleOptimalTransfer} to reduce the wavepacket dispersion. An effective beam splitter operates twice in the middle of the chain, while an effective phase shifter appears on the right half of the chain.}
  \label{f.MZ}
\end{figure}
This effect is shown in Fig.~\ref{f.MZ} for a $L{=}51$ chain where we used the
double optimal engineering scheme \cite{DoubleOptimalTransfer} to minimize the
dispersion and make $D{\simeq}1$.

\section{Conclusions}
We have proposed  a strategy to simulate arbitrary linear optics networks with minimal control on static (potentially remote) bosons in a many-site lattice. Our scheme do not require an active atomic transport with external potential, while it exploits the natural atom dynamics (the quantum walk). Ultracold atoms in optical lattices are the most natural realization in view of recent unprecedented improvements in initialization and measurements \cite{fukuhara_quantum_2013,fukuhara_microscopic_2013,GreinerNew,bakr_quantum_2009}. In particular, compared to purely photonic setups, opening up this alternative arena for linear optics can have the advantages of Fock-state (e.g. single atom) preparation (98\% fidelity \cite{GreinerNew}), naturally a very large numbers of modes (sites), and potentially a higher detection efficiency (99\% fidelity \cite{BlochNew}) 
of a single atom in comparison to a single photon. 

We have shown that fundamental operations between arbitrary remote sites are
obtained by mixing quantum walks with suitably inserted local impurities.
Several applications are considered; in particular, the achievement of a 50/50
{\it remote} beam splitter, a Mach-Zehnder interferometer, and the two-mode
Hong-Ou-Mandel effect -- rather robust against on-site interactions between
atoms till a transition from bosonic to fermionic behavior close to the
superfluid-Mott critical point. Possible sources of imperfections are discussed
in the Appendix \ref{sec:Imperfections}. 
Since atoms can be prepared and detected with high efficiency, 
our scheme may pave the way for the study of interference effects in a many-boson
optical lattice, such as the boson sampling. 
Moreover, our beam splitters, in
conjunction with on-site interactions, should open up varied possibilities for
generating non-classical states \cite{Tomaz}. 

In relation to recent experiments showing the Hong-Ou-Mandel effect in adjacent
wells \cite{kaufman}, our protocol opens up to scalable generalizations to 
many-site lattices.

\appendix
\section{Analytical derivation of the beam splitter operation for an odd chain\label{app:OddChain}}

We consider the dynamics of a single particle in a chain with $L=2N+1$ sites and
nearest neighbor interactions,
where there is an energy barrier in position $N+1$. In the single particle
sector, the Hilbert space is composed by the states $\ket n$, where $n$
specifies the position of the walking particle, and the Hamiltonian
\eqref{e.BH} can be written as 
\begin{align}
  H=-\frac{J}{2}\left(\sum_{n=1}^{2N}\ket n \bra{n{+}1} + {\rm h.c.}\right) -
  \beta J\ket N \bra
  N~. 
  \label{e.hamsingle}
\end{align}
Therefore, the reflection and transmission coefficient, as defined in section
\ref{sec:BSUniform}, are
\begin{align}
  R(t) = \bra 1 e^{-it H} \ket 1~.
  \label{e.R}
  \\
  T(t) = \bra L e^{-it H} \ket 1~.
  \label{e.T}
\end{align}
If $H=OEO^T$, in obvious matrix notation, is the eigenvalue decomposition of
the Hamiltonian, then 
$  R(t) = \sum_k O_{1k}^2 e^{-iE_kt}$, $
  T(t) = \sum_k O_{1k}O_{Lk} e^{-iE_kt}
$. 

{The Hamiltonian \eqref{e.hamsingle} is a {\it quasi-uniform} tridiagonal matrix
so the eigenvalues $E_k$ can be computed by using recurrence relations for the characteristic polynomial which in turn can be expressed as Chebyshev polynomials of first and second kind \cite{banchi_spectral_2013}.
For a odd site number chain with an impurity $\beta$ in the middle of the chain,
we find that the characteristic polynomial $\chi(\lambda)$ of the rescaled
matrix $-H/(2J)$ in terms of Chebyshev polynomials of the second kind $\mathcal{U}_n(x)$ is: 
\begin{equation}
\chi(\lambda)=(
\beta-\lambda)\mathcal{U}^{2}_{L}\left(\frac{-\lambda}{2}\right)-2\mathcal{U}_{L-1}\left(\frac{-\lambda}{2}\right)\mathcal{U}_{L}\left(\frac{-\lambda}{2}\right)~. 
\end{equation}
The latter can be expressed in terms of trigonometric function, with the substitution
$\lambda{=}-2 \cos q$ \cite{banchi_spectral_2013}. By using the formalism introduced in \cite{banchi_spectral_2013} we are able to compute the eigenvalues $E_k$ of $H$ from the roots of $\chi(\lambda)$:
We find that 
\begin{align}
  E_k &= E(q_k)~, & E(q) &= J \cos q~,
  \label{e.Eq}
\end{align}
and that there exist three types of modes. In general there are $N$ type-I modes, 
$N$ type-II modes and one out-of-band mode which has complex $q_k$ ($E(q_k)$ is real). 
As the out of band mode is localized close to the impurity we consider  only 
type-I and type-II modes in our discussion. We find
\begin{align}
  q^{\rm I}_k &= \frac{k \pi}{N+1}~ , & 
  q^{\rm II}_k &= \frac{k \pi-\phi_\beta\left(q^{\rm II}_k\right)}{N+1}~, &
  k=1,\dots,N~,
  \label{e.qIeII}
\end{align}
where 
  $\phi_\beta(q) = \arctan\left(\frac{\sin q}\beta\right)$.
Moreover, with the techniques developed in \cite{banchi_spectral_2013} one can
prove that for type-I modes $O^{\rm I}_{1k}=-O^{\rm I}_{Lk}$ while for type-II modes 
$O^{\rm II}_{1k}=O^{\rm II}_{Lk}$, so that
\begin{align}
  R(t) &\simeq U^{\rm I}(t) + U^{\rm II}(t) ~,
  &
  T(t) &\simeq -U^{\rm I}(t) + U^{\rm II}(t) ~,
  \label{e.RTU}
\end{align}
where the approximation is in neglecting the out-of-band mode, 
  $U^{\rm I,II}(t) = \sum_k \left(O_k^{\rm I,II}\right)^2 e^{-i t E(q^{\rm I,II}_k)}$,
 and 
\begin{align}
  \left(O^{\rm I}_{1k}\right)^2 &= 
  \frac{\sin^2 q^{\rm I}_k~}{N+1}~, 
  \label{e.O}
  & 
  \left(O^{\rm II}_{1k}\right)^2 &= 
  \frac{\sin^2 q^{\rm II}_k}{N+1 + \phi_\beta'\left(q^{\rm II}_k\right)}~.
\end{align}

The introduction of the energy impurity splits an incident wave packet into a
reflected and transmitted component. If the wave-function  is initially
localized in site $1$, then the transmitted wave-packet travels towards the
other boundary (site $L$) while the reflected component travels back towards the 
initial site $1$. The coefficients $|T(t)|^2$ and $|R(t)|^2$ then measures
respectively the
probability that the transmitted wave packet is reconstructed at site $L$ and 
the reflected wave-packet is reconstructed at site $1$. 
A 50/50 wave packet splitting is then realized when $|T(t^*)| = |R(t^*)|$ after
some transmission time $t^*\simeq L/J$. As we observed numerically that $|U^{\rm I}(t^*)|
\simeq|U^{\rm II}(t^*)|$, the 50/50 splitting is obtained when 
$U^{\rm I}(t^*) \simeq\pm i U^{\rm II}(t^*)$. In the following we use the
developed analytic solution to model quantitatively the wave function
splitting process.

One can show that $U^{\rm I}(t)$ is formally
analogous to one half of reflection coefficient of a chain with $N$
sites and without impurity. Exploiting this analogy,
one can use known results \cite{banchi_nonperturbative_2011} and get
\begin{align}
  U^{\rm I}(t) = 2 (-1)^{N+1}\left[\left(\frac{2N+2}{Jt}\right)^2\mathcal
    J_{2N+2}(Jt)
  -\frac1{Jt} \mathcal J_{2N+2}'(Jt)\right]~,
  \label{e.UI}
\end{align}
where $\mathcal J_n(\cdot)$ are Bessel functions of the first kind. 
From the asymptotic expansion of $U^{\rm I}(t)$ \cite{banchi_nonperturbative_2011} 
one finds a transmission
time $Jt^*\approx 2N+2+\xi(N+1)^{1/3}$, where $\xi\approx 1.019$. 
Even though the term $U^{\rm II}(t)$ is more complicated,
since we know the transfer time $t^*$ from the analysis of $U^{\rm I}$
we can find a simple expansion of $U^{\rm II}(t^*)$ in the limit $L\gg1$.
To simplify the notation,
we set $q\equiv q^{\rm II}$. By multiplying and dividing the mode expansion
 $U^{\rm II}(t) = \sum_k \left(O_k^{\rm II}\right)^2 e^{-i t E(q^{\rm
 II}_k)}$
for $e^{i(2N+2) q}$, using \eqref{e.qIeII} and going to the continuum limit 
one obtains
\begin{align}
  \label{e.U2long}
  U^{\rm II}(t^*) &= \frac1\pi \int_0^\pi dq \; 
  e^{-i t \cos q-i(2N+2)q} e^{-i2\phi_\beta(q)} \;\sin^2 q 
  \\& = \frac{(-1)^{N+1}}\pi \int_{-\frac{\pi}2}^{\frac\pi 2} dq \; 
  e^{i t \sin q-i(2N+2)q} e^{-i2\phi_\beta(q+\pi/2)} \;\cos^2 q 
  \nonumber
\end{align}
In the limit $N\gg 1$, owing to the stationary phase
approximation, the biggest contribution to the integral comes from the points
such that $q\approx \sin q$, namely $q\approx 0$. Expanding the first phase
around this point one obtains 
$ 
  t^* \sin q - (2N+2) q \approx \xi N^{1/3} q - \frac{t^*}6 q^3~.
$ 
Then, to properly take into account the scaling with $N$ we set 
$x{=} N^{1/3} \tan q$ and we perform the limit $N{\to} \infty$. The result is 
\begin{align}
  U^{\rm II}(t) &\simeq
  \frac{(-1)^{N+1}}{\pi N^{1/3}} \int_{-\infty}^{\infty} dx \; 
  e^{i\xi x - ix^3/3} f(x,\beta)~,
  \label{e.UIIinf}
\end{align}
where $f(x,\beta)$ depends also on $1/N$. Keeping only the first order, one finds
$ 
  U^{\rm II}(t^*) \simeq
  2\frac{(-1)^{N+1}}{N^{1/3}} {\rm Ai}(-\xi)
  \frac{\beta -i}{\beta +i}~,
$ 
being ${\rm Ai}(x)$ the Airy function \cite{abramowitz_handbook_1965}. 
$U^{\rm I}(t^*)$ can be obtained as well from this analysis since $U^{\rm
I}(t^*) = \lim_{\beta\to\infty} U^{\rm II}(t^*)$. 
As discussed before, a 50/50 splitter is obtained when $U^{\rm II}(t^*) = 
-iU^{\rm I}(t^*)$, i.e.  $\frac{\beta -i}{\beta +i}=-i$, and hence $\beta=1$. 
To estimate the first order corrections to this asymptotic value, we study
the first subleading order in the $1/N$ expansion. 
The result can be written in terms
of Airy functions and its derivatives and agrees with known 
asymptotic expansions of Bessel functions \cite{abramowitz_handbook_1965}.
We find that the first order correction to $\beta=1$ scales as 
 $\beta=1-\eta N^{-2/3}$, and 
 \begin{align}
   \frac{R(t^*)}{T(t^*)} = -i + \frac{i}{2} (2 \eta -\xi) N^{-2/3}
   + \mathcal O\left(N^{-4/3}\right)~.
   \label{e.RsuTinf}
 \end{align}
To have a 50/50 splitting then $\eta = \xi/2$. 
In summary, as $\xi\approx 1.019$ \cite{banchi_nonperturbative_2011}, 
the final result of this section is that for $L\gg 1$ a 50/50 splitting of the 
wave-packet is obtained when the height of the barrier is 
\begin{align}
  \label{e.bopt}
  \beta^{50/50} &\simeq 1-\frac{0.510}{N^{2/3}} \simeq 
  1-\frac{0.809}{ L^{2/3}} ~.
\end{align}
Higher order expansions can be obtained with the same method, but one has also to
consider the Euler-Maclaurin error in approximating the sum \eqref{e.RTU}
with the integral \eqref{e.UIIinf}.

\subsection{Bessel function expansion of $U^{II}(t)$}

An approximation for $U^{\rm II}(t)$ is obtained with the help of the
Jacobi-Anger expansion 
\begin{align}
  U^{\rm II}(t) = \sum_k \rho_k^{\rm II} 
  \sum_{m=0}^\infty \eta_m i^{-m}
  \mathcal J_m(t) \cos (m q_k^{\rm II}) 
  \label{e.UIIJA}
\end{align}
where $\eta_0 = 1$ and $\eta_m=2$ for $m\neq0$. As we are interested in the
neighborhood of $t=t^*$ and as the $\mathcal J_n(t)\simeq 0$ for $t\lesssim n$,
we approximate the infinite sum by only considering Bessel functions with an 
order $m\approx t^*$. By letting $m\to2N+2+m$ one can write
\begin{align}
  U^{\rm II}(t) \simeq 2(-1)^{N+1}
  \sum_{m=-M}^{M} i^{-m} c_m\;
  \mathcal J_{L+1+m}(t) 
  \label{e.UIIJAapprox}
\end{align}
where $M$ counts the number of Bessel functions considered in the approximation
and 
\begin{align}
  c_m &= \sum_k \rho_k^{\rm II} 
  \cos [(2N+2+m) q_k^{\rm II}]\\& 
    = \sum_k \rho_k^{\rm II} 
    \cos\left[m q_k^{\rm II} - 2\phi_\beta\left(q_k^{\rm II}\right)\right]~.
      \label{e.c_m}
\end{align}
We note that $c_m$ only slightly depends on the number of sites. Moreover, as the
non-uniform spacing is already included in the dependence on $\beta$ via the
function $\phi_\beta$, we consider $q$ as a continuous variable and we substitute 
$\sum_k \to \frac N \pi \int_0^\pi dq$. After some algebra, we obtain
\begin{align}
  c_m \simeq \frac 4 \pi \int_{-\infty}^{\infty} 
  \left(\frac{x-i}{x+i}\right)^m
  \frac{x^2}{\left(x^2+1\right)^3}
  \frac{x^2 \beta -2 i x+\beta}{x^2 \beta +2 i x+\beta}~dx~,
  \label{e.c_mcont}
\end{align}
valid when $N\to\infty$. 
The dependence of $c_m$ upon $\beta$ is displayed in Fig.~\ref{f.cm} for some values
of $m$. Negative values of $m$ are omitted since $c_{-m}= (-1)^m c_m$. 
For $\beta\to\infty$ only two coefficients are different from zero $c_0=\frac12$ and
$c_2=\frac14$. In that limit, with some algebra one can show that 
$U^{\rm II}(t) = U^{\rm I}(t)$: there is no transmission and the dynamics mimic
that of a chain with $N$ sites.

\begin{figure}[t]
  \centering
  \includegraphics[width=1\columnwidth]{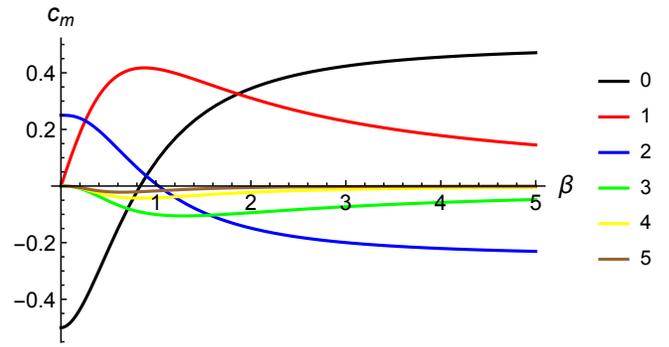}
  \caption{{ Coefficients $c_m$ vs. the impurity strength $\beta$ for
  different values of $m$}. The values are obtained from the expression 
  \eqref{e.c_mcont} valid when $N\to\infty$.}
  \label{f.cm}
\end{figure}

The dynamics of a single particle can be approximated by 
Eq.~\eqref{e.UIIJAapprox} with $M=3$. Indeed, as show in Fig.~\ref{f.cm}, the
coefficients $c_m$ are negligible for $m>3$.
By solving Eq.~\eqref{e.c_mcont} one obtains
\begin{subequations}
\begin{align}
  c_0 &=2 \beta ^2-\frac{2 \beta ^3}{\sqrt{\beta ^2+1}}-\frac{1}{2}~,
  \\ c_1 &=
  -2 \beta ^3+\frac{2 \beta ^4}{\sqrt{\beta ^2+1}}+\beta~ ,
  \\ c_2 & = 4 \beta
  ^4-\frac{4 \beta ^5+2\beta^3}{\sqrt{\beta ^2+1}}+\frac{1}{4}~,
  \\ c_3 &= -8 \beta ^5-2 \beta ^3+\frac{8
  \beta ^6+6\beta^4}{\sqrt{\beta ^2+1}}~.
\end{align}
  \label{e.c03}
\end{subequations}
This approximation reproduces the dynamics around the transfer time 
$t^*$ for chains as short as $11$ sites and becomes more accurate for
longer chains.

\section{Analytical derivation of the beam splitter operation for an even chain \label{sec:EvenChain}}
The beam splitter effect can be extended to even chains (where $L=2N$), 
by introducing an impurity coupling $J'$ in the middle of the chain. The
Hamiltonian \eqref{e.BH} in the single excitation subspace is then
\begin{align}
  H=-\left(\frac{J}{2}\sum_{n=1}^{2N-1}\ket n \bra{n{+}1} 
  +
  \frac{J'-J}2\ket N \bra {N{+}1}\right)
+ {\rm h.c.} 
  ~. 
  \label{e.hamsingle}
\end{align}
The characteristics polynomial of the rescaled matrix $-H/(2J)$ is 
\begin{equation}
\chi(\lambda)=\mathcal{U}_{2N}(-\lambda/2)+\left(1-\eta^2\right)\mathcal{U}_{N-1}^2
(-\lambda/2)~,
\end{equation}
where $\eta{=}J'/J$ and $\mathcal{U}_{n}(x)$ are Chebyshev polynomials of the second kind. Similarly to the previous case, we find that the eigenvalues of $H$ are given by 
\eqref{e.Eq}
with two different types of modes defined by 
\begin{eqnarray}
k_j^{{\rm I}}&=&\frac{j\pi -\phi^{{\rm I}}(k_j)}{N+1}~,\\
k_j^{{\rm II}}&=&\frac{j\pi -\phi^{{\rm II}}(k_j)}{N+1}~,
\end{eqnarray}
where $j\in\left\{1,\ldots, N\right\}$ and 
\begin{eqnarray}
 \phi^{{\rm I}}(k) &=& \arctan \left[ \frac{\eta \sin k}{1 - \eta \cos
 k}\right]~, \\
  \phi^{{\rm II}}(k) &=& \arctan \left[ \frac{\eta \sin k}{1 + \eta \cos
  k}\right]~.
\end{eqnarray}
We find that the type-I mode eigenvectors satisfy $O_{1 k}^{{\rm I}}{=}{-}O_{L
k}^{{\rm I}}$ while type-II modes satisfy $O_{1 k}^{{\rm II}}{=}O_{L k}^{{\rm II}}$,
similarly to the previous case, so Eqs.\eqref{e.RTU} are satisfied.
The latter expressions can be evaluated as in appendix A in the limit
$L{\gg}1$. Indeed, similarly to Eq.\eqref{e.U2long} we find 
\begin{align}
  U^{\rm I,II}(t) &\simeq
  \frac{(-1)^{N+1}}{\pi N^{1/3}} \int_{-\infty}^{\infty} dx \; 
  e^{i\xi x - ix^3/3} f^{\rm I, II}(x,\eta)~,
  \label{e.UIIinf}
\end{align}
where $x{=} N^{1/3} \tan k$ and $f^{\rm I,II}(x,\eta)$ depend also on $1/N$. Keeping only the first order, we find
\begin{eqnarray}
  U^{\rm I}(t^*) &\simeq&
  2\frac{(-1)^{N+1}}{N^{1/3}} {\rm Ai}(-\xi)\left[
  \frac{i- \eta}{i+ \eta}\right]~, \\
  U^{\rm II}(t^*) &\simeq&
  2\frac{(-1)^{N+1}}{N^{1/3}} {\rm Ai}(-\xi)
  \left[\frac{i+\eta}{i-\eta}\right]~,
\end{eqnarray}
being ${\rm Ai}(x)$ the Airy function \cite{abramowitz_handbook_1965}. 
As discussed before, the 50/50 beam splitting  condition is achieved for $U^{\rm I}(t^*) {=} 
iU^{\rm II}(t^*)$ that lead to the coupling value $\eta{=}\sqrt{2}{-}1$.
Deviations from this value due to finite size  effects have been investigated numerically, as shown in Fig. \ref{fig:EvenUniform}.

\section{Imperfections \label{sec:Imperfections}}
We discuss how possible imperfections in real experiments could affect our
theoretical results. 
We focus on beam splitters and unmodulated chains.  
Firstly we consider a non perfectly localized optical impurity with a 
Gaussian profile 
$\beta_j{=}\beta^{50/50} \exp [{-}(N{+}1{-}j)^2/\sigma^2]$. 
We find that 
$\epsilon{=}(\vert R(t^*)\vert {-} \vert T(t^*) \vert){/}\vert R(t^*)\vert{\simeq}0$ 
for ${\rm FWHM}{\lesssim}0.66 a$, being $a$ the lattice spacing. 
With suitably changed $\beta^{50/50}$ we observe 
small deviations ($\epsilon{\lesssim} 5\%$) 
when ${\rm FWHM}{\lesssim}8a$
($L{=}51$).
An open-ended chain can be realized in an extended lattice by
adding two strong localized fields $\beta_{\rm walls}$ on sites
$0$ and $L{+}1$. When $\beta_{\rm walls}$ is sufficiently high, the 
effective chain resulting between sites $1$ and $L$ 
is almost decoupled from the rest of the lattice, {\it e.g.}
$\epsilon{<}5\%$ 
when $\beta_{\rm walls}{\gtrsim} 3$. 
A non zero curvature of the optical lattice can be modeled via 
an effective chemical potential 
\cite{wurtz_experimental_2009} 
$\mu_j^{\rm eff}{=}\mu {-} J\omega^2 j^2/2$. 
When $\omega {\lesssim} 0.03$ results do not  deviate from the ideal case.

\begin{figure}[t]
  \centering
  \includegraphics[width=1\columnwidth]{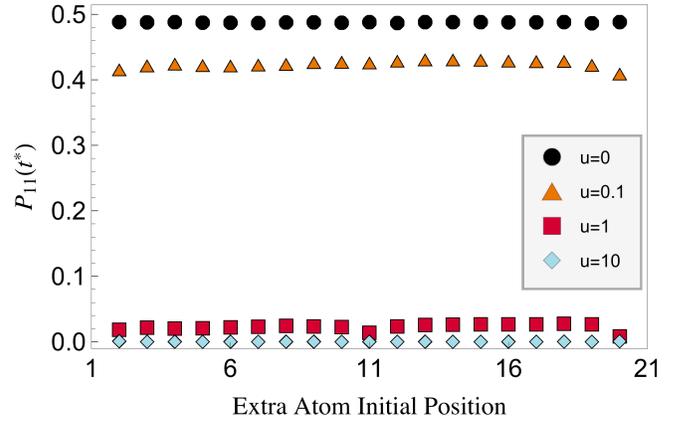}
  \caption{ Left bunching probability at the transmission time $t^*$ 
    for different values of $u=U/J$, when there is another particle in the
  lattice. The initial state is $\ket{\psi(0)}{=}a_1^\dagger a_{21}^\dagger a_j^\dagger\ket{0}$ where $j\in\left\{ 2,\ldots,20\right\}$
  is the position of the unwanted particle that is represented by the $x$-axis position. The chain has $L{=}21$ and it is minimal engineered with the double optimal coupling scheme. 
 }
  \label{fig:Impurity}
\end{figure}

Finally, 
in Fig.~\ref{fig:Impurity} we study the effect of an unwanted extra particle in our scheme, 
that might be present in the chain because of an imperfect initial state preparation. We consider the initial state
\begin{align}
  \ket{\psi(0)} &= a_1^\dagger a_L^\dagger a_m^\dagger \ket 0, &
  m=2,\dots,L-1~,
  \label{TrePart}
\end{align}
and we study the probability to find two particles in the first (last) site of the chain at the transmission time $t^*$ as a function of $u$ and the initial position of the unwanted particle, irrespectively of the final position of the third particle. Defining $\ket{\phi_j}{=}a_1^\dagger a_1^\dagger a_j^\dagger \ket{0}$ the final state of the system at time $t^*$, in Fig. \ref{fig:Impurity}, we plot the quantity 
\begin{equation}
P_{11}(t^*,m)=\sum\limits_{j=1}^{L}\vert  \langle  \phi_j\vert U(t^*)\vert \psi(0)\rangle    \vert^2
\end{equation}
as a function of the initial position of the third particle $m$, for a $L{=}21$ chain and . 
We see that for reasonably small values of $u$ the 
unwanted particle introduces no observable effects independently from its initial position.

\section*{Acknowledgements}
The research leading to these results has received funding from the European Research Council under the European Union's Seventh Framework Programme (FP/2007-2013) / ERC Grant Agreement n. 308253.

\end{document}